\documentclass[12pt,preprint]{aastex}
\usepackage{stmaryrd}
\usepackage{txfonts}
\usepackage{mathrsfs}
\usepackage{graphicx}

\newcommand{\beq}{\begin{equation}}
\newcommand{\eeq}{\end{equation}}
\newcommand{\bea}{\begin{array}}
\newcommand{\eea}{\end{array}}

\shorttitle{Dusty Disks around White Dwarfs} \shortauthors{Dong
{\it et al.} }

\begin{document}

\title{Dusty Disks around White Dwarfs I: Origin of Debris Disks}

\author{Ruobing Dong$^{1,2}$\altaffilmark{*}, Yan Wang$^{1,3}$,
D. N.C. Lin$^{1,4}$, and X.-W. Liu$^{1,5}$}

\affil{$^1$Kavli Institute of Astronomy and Astrophysics,
Peking University, Beijing, China}
\affil{$^2$Department of Astrophysical Sciences, Princeton
University, Princeton, NJ 08544, USA; rdong@astro.princeton.edu}
\affil{$^3$Department of Astronomy and Astrophysics , Penn State
University, State College, PA 16802, USA; yuw123@psu.edu}
\affil{$^4$Department of Astronomy and Astrophysics, University of
California, Santa Cruz, CA 95064, USA; lin@ucolick.org}
\affil{$^5$Department of Astronomy, Peking University, Beijing
China; liuxw@bac.pku.edu.cn}
\altaffiltext{*}{Correspondence should be addressed to
rdong@astro.princeton.edu.}

\begin{abstract}
A significant fraction of the mature FGK stars have cool dusty disks
at least an orders of magnitudes brighter than the solar system's
outer zodiacal light. Since such dusts must be continually
replenished, they are generally assumed to be the collisional
fragments of residual planetesimals analogous to the Kuiper Belt
objects. At least 10\% of solar type stars also bear gas giant
planets. The fraction of stars with known gas giants or detectable
debris disks (or both) appears to increase with the stellar
mass. Here, we examine the dynamical evolution of systems of
long-period gas giant planets and residual planetesimals as their host
stars evolve off the main sequence, lose mass, and form planetary
nebula around remnant white dwarf cores. The orbits of distant gas
giant planets and super-km-size planetesimals expand adiabatically.
During the most intense AGB mass loss phase, sub-meter-size particles migrate toward
their host stars due to the strong hydrodynamical drag by the
intense stellar wind.  Along their migration paths, gas giant planets
capture and sweep up sub-km-size planetesimals onto their mean-motion
resonances. These planetesimals also acquire modest eccentricities
which are determined by the mass of the perturbing
planets, the rate and speed of stellar mass loss. The swept-up
planetesimals undergo disruptive collisions which lead to the
production of grains with an extended size range. The radiation drag
on these particles is ineffective against the planets' resonant
barrier and they form 30-to-150-AU-sizes rings which can effective
reprocess the stellar irradiation in the form of FIR continuum. We
identify the recently discovered dust ring around the white dwarf WD
2226-210 at the center of the Helix nebula as a prototype of such
disks and suggest such rings may be common.
\end{abstract}


\section{Introduction}
The discovery of over 300 planets shows that at least 10 \% of
nearby solar type stars have Jupiter-mass planets around them
(Cumming {\it et al.} 2008). According to widely adopted sequential
accretion hypothesis, these planets formed through condensation
of heavy elements, emergence of planetesimals which coagulated
into protoplanetary embryos and cores prior to efficient gas
accretion. Through a series of population-synthesis simulations
based on this scenario (Ida \& Lin 2004a,b, 2005, 2008a,b), we
were able to reproduce the observed mass-period ($M_p-P$)
distribution of the known gas giant planets (Schlaufman
{\it et al.} 2008) around a modest fraction ($\eta_J$) of FGK
main sequence dwarf stars.

Mass accretion rate $\dot M$ from protostellar disks onto stars more
massive than the Sun is observed to increase with square of the stellar
mass {\it i.e} $M_\ast^2$ (Natta {\it et al.} 2006, Garcia-Lopez
{\it et al.} 2006, cf Clarke \&  Pringle 2006). If the initial mass of
these disks also increases with  $M_\ast$, the greater availability
of heavy elements would provide more rapid growth and larger of
protoplanetary embryos around intermediate mass stars (Ida \& Lin 2004b,
2005, Kennedy {\it et al.} 2007). Furthermore, the extended
local surface density and pressure maxima at the ionization radius
(where the mid-plane temperature $T_c \simeq 1000$ K), snow line
(Kretke {\it et al.}  2009), and outer edge of the dead zone provide
more efficient barrier against hydrodynamic drag for the sub-km
grains and type-I migration for the critical-mass ($\sim 10 M_\oplus$)
cores.  The enhance retention efficiency of these building blocks
enhance emergence probability of gas giants (Laughlin {\it et al.}
2004, Ida \& Lin 2004b).

On the observational side, high-precision radial-velocity surveys of
intermediate-mass main sequence stars have been handicapped by their
hot and active atmosphere as well as their fast spin (Griffin {\it et
al.} 2000). However, these problems are significantly reduced when
these stars evolve off their main sequence. Several planets have been
found around G sub-giants and giants (Sato {\it et al.}  2005) which
have more expanded envelopes and relatively cool atmospheres.
Preliminary surveys indicate that the fraction of these relatively
massive G giant stars with gas giant planets ($\eta_J \sim 1/3$) may
be more than double that around the solar-type G dwarf main sequence
stars (Sato {\it et al.} 2007, Johnson {\it et al.} 2007, Dollinger
{\it et al.}  2007, Hatzes 2008, Johnaon 1008).

Another area where observations provide valuable clues on the planet
formation process is the search for residual building-block
planetesimals around their host stars. Since 1995 a population of
Kuiper Belt objects (KBO's) has been found in the solar system and
beyond the orbit of Neptune (Jewitt \& Luu 1995, Brown {\it et al.}
2005). These objects are thought to be the parent bodies of a
diffuse dusty ring which emits cold zodiacal light with far infrared
(FIR) radiation.

A series of recent systematic 70 $\mu$m (FIR) observational surveys
with the Spitzer Infrared Space Telescope indicate that 1) none of
the nearby M stars show detectable excess (Gautier {\it et al.}
2007), 2) 15 \% of 274 FGK stars show excess (Bryden {\it et al.}
2006, Beichman {\it et al.} 2006, Trilling {\it et al.} 2008), 3) 30
\% of 160 A stars also show excess (Rieke {\it et al.} 2005; Su \&
Rieke 2007). The threshold level for a marginal detection by the
Spitzer Telescope is at least an order of magnitude more intense
that of the solar system zodiacal light.

Since both the fraction of stars with gas giants ($\eta_J$) and that
with detectable debris disks ($\eta_d$) increases with the stellar
mass ($M_\ast$), we can extrapolate that a major fraction of
intermediate-mass stars may have both gas giants and persistent,
rich debris disks.  This inferred association of gas giant planets
with debris disks is highlighted by the recent discovery (with
adaptive imaging technique) of three gas giants and a debris disk
outside them around an A5V star HR 8799 (Marois {\it et al.} 2008).
The mass and periods of these three planets are (10, 10, 7) $M_J$
and (24, 38, 68) AU.  Around this $M_\ast= 1.5 M_\odot$ star, the
ice line is located at $\sim$10 AU. In \S2, we provide a more
detailed discussion on the association of debris disks and planet
formation in the Solar System and around nearby stars.

With the evidence of existence of many detectable debris disks and
gas giant planets around nearby, mature, intermediate-mass stars, we
now consider their dynamical evolution during the post main sequence
evolution of their host stars. The goal of our study here is to
identify whether the inconspicuous planet building blocks may reveal
their presence. The analysis to be presented below is particularly
relevant for the AGB mass loss phase where the mass loss rates range up to $\sim 10^{-4} M_\odot$ yr$^{-1}$ and where most of the stars may reach this mass loss rate near the end of the AGB phase (Willson 2000). The typical outflow
speed is $\sim 10$ km s$^{-1}$ and $\dot M_\ast$ may vary on the
time scale of $10^{3-4}$ yr (Weidemann 1987, Blocker 1997). After
losing mass on the AGB, these stars rapidly evolve across the HR
diagram to form planetary nebula around emerging white dwarfs.
During the post AGB phase, although high-speed mass loss continue
to occur, the mass flux is much reduced (cf Kwok 1982).

In this paper, we mostly consider stars with a main-sequence masses in
the range of $2 M_\odot < M_\ast < 4 M_\odot$ (A-F stars). We choose
these stars not only because many, perhaps most of them, have gas
giants and debris disks (similar to those around the less massive Sun
and HR 8799), but they are also thought to be the progenitors of some
well known planetary nebulae such as the Helix nebula where the mass
of the remnant white dwarf WD 2226-210 is $\sim 0.58 M_\odot$ (Su {\it
et al.} 2007). This rapid and large-magnitude change in the
gravitational potential can lead to great orbital expansions for their
planets. This evolutionary tendency may lead to the possibility of
resolvable HST imaging of long-period planets which are otherwise
undetectable.

Evidence for the existence of residual planetesimals comes from the
recent discovery of a dusty ring around a young white dwarf WD
2226-210 at the center of the Helix nebula (Su {\it et al.}  2007).
This ring extends between about 35 and 150 AU from a central white
dwarf WD 2226-210 (well interior to the complex, helix structure),
and have a total mass of about 0.13 $M_ \varoplus $. We propose that
this ring is the byproduct of planets and planetesimals' orbital
evolution during the epoch when the central star rapidly lost most
of its mass. Based on the extrapolation of the system around HR
8799, our basic conjectures are:

\noindent
1) gas giant planets and debris disks are common around A main sequence stars;

\noindent
2) at the epoch of their main sequence turnoff, these relatively
massive stars have debris disks with more than 10$M_\oplus$ in total
mass and most of it is contained in km-size residual planetesimals

\noindent
3) gas giant planets' orbits expanded adiabatically by
nearly an order of magnitude while there host stars evolve through
AGB mass loss phase with rapid stellar mass loss;

\noindent
4) the KBO-equivalent planetesimals are captured onto the main motion
resonance of gas giants; and

\noindent
5) through collisions, these parent bodies generate dust which
reprocess the radiation of their central star.

These basic assumptions are based on their solar system and HR 8799
analogues. A detailed and systematic analysis of these processes is
useful for understanding the long-term evolution of this class of
planetary systems. In \S3, we first construct a working model for a
solar-system like configuration. The orbital evolution of the
planets and planetesimals is computed and analyzed in \S4. Here we
take into account of the central star's mass loss, the hydrodynamic
drag by the expanding envelope and the planet's dynamical
perturbation on the residual planetesimals. In \S5, we show that
collisions between the resonant planetesimals lead to their
disruptions and fragmentation. We derive the condition for the
retention of dust grains into rings outside the orbits of some gas
giant planets. We model the structure of the disk around WD 2226-210
in \S6.  Finally, we summarize our results and discuss their
implication, especially in the context of possible detectable gas
giant planets around young white dwarfs in \S7.

\section{Association of debris disks and gas giants}

In this section, we briefly recapitulate existing observational
information on the inventory of residual planetesimals in the outer
solar system and around mature, intermediate-mass main-sequence stars.
These data will be used in the subsequent sections for the development
of our working model.

\subsection{Inventory of residual KBO's in the outer solar system.}
KBO's are marginally preserved at their remote location because
planetary perturbation (by gas and ice giants) is relatively weak
(Duncan \& Quinn 1993). However, their present-day spacial
distribution also poses a challenge to the formation of the largest
KBO's which have masses comparable to Pluto. If their building blocks
were initially smoothly distributed with respect to their distances
from the Sun, {\it in situ} formation of such large KBO's would
require more than a few $M_\oplus$ of planetesimals in the region
30-50 AUs (Stern 1996, Youdin \& Chiang 2004). This required total
mass is $\sim 2$ orders of magnitude larger than the total reservoir
of all known KBO's. Dynamical perturbation by Uranus and Neptune can
induce a substantial depletion from a massive initial inventory to the
present-day KBO reservoir (Duncan \& Quinn 1993). This stringent
requirement would be relaxed under the assumption that the most
massive KBO's were formed closer to the Sun and then scattered to
their present-day location by gas or ice giant planets either during
the advanced phase of their formation (Zhou \& Lin 2007) or through
long term dynamical evolution of the system (Duncan {\it et al.}
1987).

It is possible that the size distribution of the KBO's extend well
below the present detection limit. Elsewhere in the solar system,
the crater density on the unprocessed moons of the outer planets
(Hartmann 1999) indicate that they have been bombarded by meteorites
with a size distribution $dN/ds \propto s^{-3.5}$ as observed among
the asteroids (Gradie {\it et al.} 1989). This distribution is
probably the byproducts of a collisional equilibrium (Donanyi 1969,
Williams \& Wetherill 1994, Farinella \& Davis 1996, Tanaka \& Ida
1996, Kenyon \& Bromley 2002, Weidenschilling 2004). On the
small-size limit, particles in Saturn's rings (Cuzzi {\it et al.}
1980) and grains in the interstellar medium (Mathis {\it et al.}
1977, Laor \& Draine 1993) follow a similar pattern. With such a
size distribution, most of the surface area is covered by the small
particles while a large fraction of the total mass is contained in
the large particles.

As a population, the KBO's have a directly-observed size
distribution $dN/ds \propto s^{-5}$ for $s>100$km (Trujillo {\it et
al.} 2001) whereas, below $s \sim 70 $ km, $dN/ds$ is considerably
shallower than $s^{-3.5}$, ({\it i.e.} there is a deficit of small
particles)\ down to the detection limit of a few $10$ km (Bernstein
{\it et al.} 2004). This break in the slope of the size distribution
implies a large fraction of the KBO's total mass is contained in
particles with $s \sim 10-100$ km. One possible reason for this
complex power-law distribution is the collisional fragmentation of
KBO's as piles of loose gravels (Pen \& Sari 2005).

At their present distance, it is difficult to directly image
planetesimals smaller than a few $10$km. However, micron to sub-mm
particles in the asteroid belt and Kuiper belt scatter and reprocess
the solar radiation to produce warm and cold zodiacal light (Backman
{\it et al.} 1995). Due to the Poynting-Robertson (PR) drag induced on
them by the stellar photons, the grains undergo orbital decay on
time scale much shorter than the age of the Sun. Thus, these grains
must be continually replenished, most likely through the break up of
their parent bodies in the Kuiper Belt (Kenyon \& Luu 1999).
Energetic collisions can indeed lead to the disintegration of parent
bodies and a catastrophic production of very small fragments
(Fujiwara {\it et al.} 1977, Petit \& Farinella 1993, Benz \&
Asphaug 1999). The KBO's can provide sufficiently frequent
collisions to replenish the supply of small grains needed to account
for the observed level of the zodiacal light provided 1) a
non-negligible fraction of their total mass is contained in KBO's
with $< 1-100$ km, and 2) a modest fraction of these collisions
actually lead to catastrophic fragmentation.

Asymmetric reradiation of stellar light by spinning bodies can lead
to a torque on their orbits.  This effect is commonly known as the
Yorkovski (YORP) effect (Bottke {\it et al.}  2000, de Pater \&
Lissauer 2001). The characteristic YORP-induced orbital-evolution
timescale for planetesimals with a density $\rho_p$, semi major axis
$a_i$, and Keplerian speed $v_K = (GM_\ast/a_i)^{1/2}$ is
\begin{equation}
\tau_Y \sim \pi c \rho_p s a_i^2 v_K/f L_\ast \label{eq:tyorp}
\end{equation}
where $M_\ast$ and $L_\ast$ are the mass and luminosity of the host
star. The magnitude of the efficiency factor $f$ is determined by
the fractional amplitude of seasonal and diurnal temperature
variations. The orbits of planetesimals with prograde spins expand
as they experience a positive force (and $f$) whereas planetesimals
with retrograde spins undergo orbital decay with a negative $f$. The
YORP effect enhances the PR drag for small particles. It can also
lead to significant diffusion in the spacial distribution of km-size
planetesimals during the main sequence life span of intermediate-mass
stars. At 10 AU from stars with $L_\ast$ ten times larger than that of
the Sun, meter-size particles may undergo orbital decay within a
Myr (see \S\ref{sec:model} below).

\subsection{Debris disks around nearby intermediate-mass stars}

In the quest to determine the properties and understand the origin of
planets around other stars, it is often useful to identify their solar
system analogs. Although it is impossible to directly detect residual
planetesimals similar to the KBO's, the counterpart to the solar
system zodiacal light have been found with Spitzer Space Telescope in
the form or MIR and FIR excesses around a significant fraction of
nearby mature solar type stars (Rieke {\it et al.} 2004, Beichman {\it
et al.} 2005, Moro-Martin {\it et al.} 2007, Hillenbrandt {\it et al.}
2008). Such excesses are assumed to be due to the reprocessed stellar
radiation by debris dusty disks or rings at $\sim 2-20$ AU from their
host stars (Bryden {\it et al.} 2006, 2007, Trilling {\it et al.}
2008). The relevant-size grains are either blown away by radiation
pressure or spiral toward their host stars due to radiation drag (both
PR and YORP effects) on time scale much shorter than the age of the
system (Artymowicz \& Clampin 1997). They must be continually
replenished by the collisions between relic planetesimals, analogous
to the those in the Kuiper Belt and Oort Cloud (Kenyon \& Bromly 2002,
2004, Wyatt {\it et al.} 2007, Moro-Martin {\it et al.}  2007, 2008).

The Spitzer Space Telescope's detection threshold for these system
is at levels at least an order of magnitude brighter than the outer
solar Zodiacal light. It is entirely possible that all stars may
have debris disks similar to that around the Sun. Some known debris
disks have FIR excesses are $10^{1-3}$ times larger than that
associated with the zodiacal light in the solar system (Bryden {\it
et al.} 2006). On average, the rate of collisional dust production
is approximately proportional to the square of the surface density
of their parent bodies. Presumably, stars with detected debris disks
have richer reservoirs of relic planetesimals than the sun, albeit
occasional major impacts may lead to brief episodic flare ups in the
IR excess around individual systems (Kenyon \& Bromly 2004,
Grigorieva {\it et al.} 2007).

In the attempt to differentiate transient events and average properties,
it may be fruitful to observationally determine the fraction of stars
with detectable debris disks in a given population-analysis sample.
In \S1, we have already cited observational indications that
$\eta_d$ is an increasing function of $M_\ast$. This correlation of
debris disks with the stellar mass can be partially attributed to
observational selection effect because 1) A main sequence stars are
generally younger and 2) their disks with similar fractional
luminosity, relative to theirs, are brighter, warmer, and easier to
detect than those around FGK dwarfs (Bryden, private communication).
Nevertheless, this correlation is consistent with the expectation
inferred from 1) the fraction of stars with planets, and 2) the disk
accretion rate and mass around 1-3 Myr old stars are increasing
functions of the stellar mass.

Further evidences for the collisional origin of debris-disk dust
particles can be inferred from the general decline in the intensity
of the debris disks with the age of individual host stars (Rieke
 {\it et al.}  2004). As a population, the fraction of A stars with
70 $\mu$m excess decreases from 48 \% around those younger than 30
Myr to 12 \% for those older than 400 Myr. These signatures are
consistent with the expectation that both dynamical and collisional
processes lead to the depletion of these parent bodies.
Nevertheless, a fraction of relatively massive A stars returns
detectable debris disks throughout their main sequence lifespan.

The fraction planet-bearing FGK main sequence stars with FIR
signature appears to be comparable to those without any known
Jupiter-mass planets. After correcting some unfavorable
observational selection effects, the intensity of the FIR excess for gas
giant planet-bearing stars is $\sim 3$ times that for stars without
any planets, albeit the statistical significance of this correlation
is weak (Bryden {\it et al.} 2007). In contrast to the strong
correlation between the metallicity of FGK dwarfs and their
planet-bearing probability (Fischer \& Valenti 2005, Santos {\it  et
al.} 2005), there appears to be no correlation between it and
detectable debris disks (Greaves {\it et al.} 2007). This dichotomy
is probably due to the retention of debris-disk parent bodies
depends much less sensitively on the surface density of
planet-building blocks in their nascent disks (Wyatt {\it et al.}
2007) than the emergence of gas giant planets (Ida \& Lin 2008a, b).
However, the relatively high fraction of stars with planets around
more massive (with mass $M_\ast > 1-2 M_\odot$) stars is independent
of their metallicity (Pasquini {\it et al.} 2007). We have already
indicated above that planet formation around the relatively massive
stars may be more prolific not only because the protostellar disks
around them have greater inventories of planet building blocks but
also that the planetesimals formed more readily and are better
retained in these disks.

Since both $\eta_J$ and $\eta_d$ increases with $M_\ast$ and attain
modest values ($0.2-0.3$) for intermediate-mass stars, a significant
fraction of them (such as HR 8799 and Fomalhaut Kalas {\it et al.}
2008) may contain both. Our objective in this paper is to
consider the dynamics of the planets and residual planetesimals
during the post main sequence evolution of some intermediate-mass
host stars.

\section{A working model}
\label{sec:model}
Dynamical response of planets as a consequence of mass loss during
the planetary nebula stage has been considered by several authors in
the past mostly in the context of metal-rich DZ white dwarfs (Stern
 {\it et al.} 1990, Sackmann {\it et al.} 1993, Parriott \& Alcock
1998, Siess \& Livio 1999a, b, Debes \& Sigurdsson 2002, Jura 2006).

For the origin of dusty disks in planetary nebulae, we consider
similar effects, with a greater detailed treatment of both orbital
dynamics and gas drag effects. We idealize the dynamics of the
system into a series of interaction between three populations of
objects: planetesimals with mass $m_j$, planets with mass $M_j$, and
a central star with mass $M_\ast$.

There are two mean factors which influence the orbital change of a
planet around the central star during the epoch of its mass loss:
the change of the star's mass, which reduces its gravity and causes
the planet's orbit to expand; and the gas drag by the outflow, which
induces orbital decay. But the magnitude of this drag on a gas giant
planet is negligible so that the equation of motion for the gas
giant is reduced to the standard Kepler's problem.

But the drag effect has a significant effect on the orbit of the
planetesimal. In addition, it is also perturbed by the gravity of
the planet. In a non rotating frame centered on the star, the
equation of motion of a planetesimal at a position ${\bf r}_i$ can
be expressed as
\begin{equation}
{d^2 {\bf r}_i \over d t^2} = - {G M_\ast {\bf r}_i \over a_i^3} -
\Sigma_j {G M_j ({\bf r}_i - {\bf r}_j) \over \vert {\bf r}_i - {\bf
r_j}\vert^3} - \Sigma_j {G M_j {\bf r}_j \over r_j^3} - {{\bf f}_i
\over m_i } \label{eq:motion}
\end{equation}
where ${\bf r}_j$ is the position of the planet and $M_\ast$ is a
function of time.  On the right hand side of the above equation,
gravity of the star and planet are expressed in the first two terms.
The third term corresponds to the indirect force due to the rotation
of the frame (Murray \& Dermott 1999) and the forth term represents
the gas drag force on the particles.

In the Stoke's region, the magnitude of the drag force on a
planetesimal by the wind of its host star can be expressed as
\begin{equation}
{\bf f}_i=\pi \rho_{wind} (s^2+s_g^2) ({\bf {\dot r}}_i - {\bf
v}_{wind}) \vert {\bf {\dot r}}_i - {\bf v}_{wind} \vert
\label{eq:fdrag}
\end{equation}
where $v_{wind}$ is the wind velocity, $s = (3 m_i / 4 \pi
\rho_p)^{1/3}$ and $s_g = G m_i/v_p^2$ are the physical and
gravitational radius of a planetesimal with a mass $m$ and density
$\rho_p$ (cf Supulver \& Lin 2000). In the radial direction,
this drag contribution is much
smaller than the gravity from all three bodies. However, in the
azimuthal direction, the drag slows down the planetesimals' initial
Keplerian orbits (with a velocity $v_K = (G M_\ast/r)^{1/2}$ at a
distance of $r$) to migrate inward on a characteristic time scale

\begin{equation}
\tau_{\rm drag} = \left| {r \over {\dot r}}\right| = \left| {v_K
\over 2 {\dot v}_K}\right| = {v_K m_i \over 2 f_i} \simeq {8 \pi
\rho_p s r^2 \over 3 \dot{M}_\ast} {v_{wind} \over v_K}.
\label{eq:taudrag}
\end{equation}\

In the above expression, we have neglected the gravitational focussing
effect (which is small for km-size planetesimals) and assumed a steady
spherically symmetric wind such that the background gas is determined
by the stellar mass loss rate
\begin{equation}
\dot{M}_\ast=4\pi r^2 \rho_{wind} v_{wind}
\end{equation}
and the wind speed $v_{wind}$ is generally assumed to be 10km/s
which is larger than $v_K$ for $r$ greater than a few AU.

We can include other effects to the equations of motion. For
example, the shape of some planetary nebulae significantly departs
from spherical symmetry (O'Dell {\it et al.} 2004). In a follow up
paper, we add a $J_2$ component to the stellar gravity. It is also
possible to include the gravity of additional stellar or planetary
companions. In the current investigation, we neglect the PR and YORP
effects during the AGB mass loss phase. Based on eq(\ref{eq:tyorp}),
we estimate that these effects can lead to significant orbital
evolution during the brief AGB mass-loss phase (for $\sim 10^4$ yr) only
for sub-meter-size particles. Comparing eqs (\ref{eq:tyorp}) and
(\ref{eq:taudrag}), we find that the hydrodynamic drag has a
stronger impact on the orbital evolution of all particles at all
locations if
\begin{equation}
{\dot M} > {\dot M_Y} \equiv 8 f L_\ast v_{wind} / c v_K^2 \sim
10^{-6} M_\odot {\rm yr}^{-1} \label{eq:mdotyorp}
\end{equation}
during the AGB mass loss phase.

However, during the main sequence life span $\tau_\ast$ of A stars,
YORP can induce residual planetesimals with sizes smaller than
\begin{equation}
s_Y = \tau_\ast f L_\ast/(\pi c \rho_p a_i^2 v_K) \sim (a_i /10
{AU}) ^{3/2} {\rm km}
\end{equation}
to migrate over a significant fraction of their $a_i$.  The inward
migration of planetesimals may be halted after they are captured by
the mean motion resonances of any embedded gas giants whereas the
outward migration is limited to be less than $10^2$ AU (see further
discussions on the initial planetesimal reservoir below).

In order to fully explore possible range of initial conditions, we
solve the equation of motion with a Hermit scheme which is kindly
provided by Dr Sverre Aarseth (2003).

\section{Orbital evolution of planets and planetesimals due stellar mass
loss}

We recapitulate here the response of planets and planetesimals due to
the changing gravitational potential during the rapid loss phase of
their host AGB or post-AGB stars.

\subsection{Adiabatic orbital evolution of planets and large
planetesimals}

In our model, there are two main factors which affects the planet's
orbital evolution: the reduction of the central mass and the gas
resistance. The resistance term increases with $s^2$ while the
gravitational term increases with $m_i \sim s^3$. For relatively
large planetesimals, gas resistance is unimportant and its orbital
change is determined only by the central mass. Provided the time
scale of the host star's mass loss $\tau_M = \left| M_\ast / {\dot
M}_\ast \right|$ is long compared with their Keplerian orbital time
scale $\tau_K = (a_i^3 / GM_\ast)^{1/2}$, the planetesimals' angular
momentum $h=vR$ does not change during this episode of mass loss,
{\it i.e.} it is an adiabatic invariant. The planetesimal also
retain their radial action so that a centrifugal balance is
maintained in which $v^2=GM_\ast/a_i$. Consequently, $a_i = h^2 /
GM$ and the orbit expands as the host star loses mass (Duncan \&
Lissauer 1998).

In contrast, distant planets and planetesimals with $\tau_K >
\tau_M$ react to the impulsive change in the gravitational potential
such that they would escape if their host stars lose more than half
of their mass. During the AGB mass loss phase, the characteristic
time scale for mass loss, $\tau_M \sim 10^{4-5}$ yr.  All
planets and planetesimals with period less than $\tau_M$ (or semi
major axis within several $100$ AU) undergo orbital
expansion rather than escape.

The characteristic orbital expansion time scale is
\begin{equation}
\tau_{\rm exp} = \left|{a_i \over {\dot a}_i}\right| = \left|{M_\ast
\over {\dot M}_\ast}\right| = \tau_M. \label{eq:tauexp}
\end{equation}
From equations (\ref{eq:taudrag}) and (\ref{eq:tauexp}), we find that
the magnitude of $\tau_{\rm exp} \simeq \tau_{\rm drag}$ for particles
with sizes
\begin{equation}
s \sim s_c \equiv {3 M_\ast \over 8 \pi \rho_p a_i^2} \left( {v_K \over
v_{wind}} \right) \simeq 0.1 \left( M_\ast \over M_\odot \right)^
{1.5} \left( 1 {g \ cm^{-3}} \over \rho_p \right) \left(a_i \over 10
{\rm AU} \right)^{-2.5} \left( 10 {km \ s^{-1}} \over
v_{wind} \right) {\rm km} .
\label{eq:sc}
\end{equation}
The above equation is applicable in the $\rho_{wind} \propto 1/r^2$ (or $v_{wind}$~=~constant) region of the stellar wind, and with the planet orbital radius greater than a few stellar radii. During the AGB mass loss phase, orbits of planetesimals with $s > s_c$
expand during the stellar mass loss. Orbits of planetesimals with $s < <
s_c$ undergo orbital decay on a time scale
\begin{equation}
\tau_{\rm drag} \sim (s/s_c) \tau_M.
\label{eq:taudrag2}
\end{equation}

\begin{figure}
  \includegraphics[angle=-90, width=150mm] {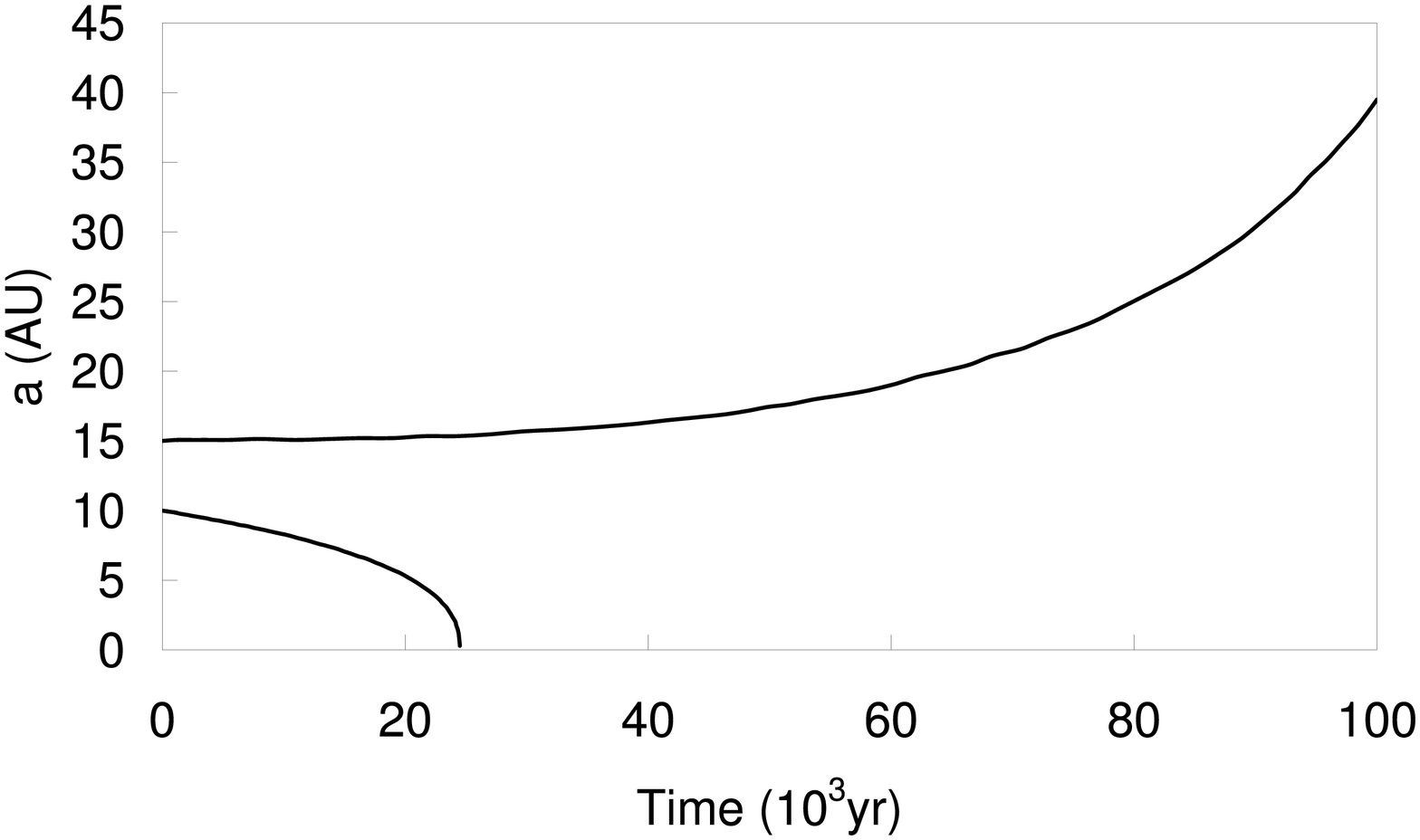}\\
  \caption{Orbital evolution of two planetesimals around a mass-losing star.
  The initial mass of the central star in this simulation is 4 $M_\odot$.
  It loses $75\%$ of its mass at a constant mass loss rate over $10^5$
  years (from now on it is called "linear mass loss"). The upper and lower curves in the plot show the
  evolution of semi-major axis verse time of two identical planetesimal, with
  density $\rho = 3$ g/cm$^3$ and size $s=1$ km. The upper one starts from 15 AU,
  and its size is larger than the $s_c$ at that position, so its orbit expands.
  While the lower one starts from 10 AU, with its size being smaller than $s_c$
  at that position, it undergoes orbital decay, and finally drops into the star
  within about 25,000 years.}
  \label{fi:sc}
\end{figure}

The relation of $s_c$ with orbital radius is showed in Figure
\ref{fi:sc} which indicates a km-size planetesimal would be larger
than $s_c$ at $r=15$ AU from a mass-losing star so that its orbit
would expands whereas it would be smaller than $s_c$ at $r=10$ AU so
that it would undergo orbital decay during the mass loss. This
dependence arises because the gas density in a steady wind decreases
with $a_i$. All planetesimals with sizes larger than $s_c$ at their
initial location are expected to expand, albeit with size-dependent
speeds.

We are primarily interested in Kuiper-Belt equivalent regions
because 1) that is the region where FIR from the debris disks
implies the potential presence of a large population of residual
planetesimals, 2) it is outside the region where gas giants are
expected to form, and 3) it is also outside the quasi hydrostatic
atmosphere of AGB's so that planetesimals may survive the evolution
of their host stars prior to the AGB mass loss phase. For this
region, $s_c \sim 1$ km. In \S2, we have already indicated that a
significant fraction of the total mass in the Kuiper Belt may be
contained in KBO's in this size range. In the above section, we also
indicated that due to the YORP effect during the mean sequence
evolution of their central stars, km-size KBO's (with $s < s_Y$)
have a tendency congregate near the outer mean motion resonances of
long-period gas giants.

Around any given host star, the mass loss rate during the AGB mass
loss phase may have large amplitude fluctuations. Since both the
orbit expansion in response to the decreasing stellar mass and the gas
drag in the stellar wind are proportional to the stellar mass loss
rate, $s_c$ is independent of $\dot M$ [note that $s_c$ is
defined by $\tau_{\rm exp} \simeq \tau_{\rm drag}$ and both
$\tau_{\rm exp}$ and $\tau_{\rm drag}$ are inversely proportional to
$\dot M$]. Nevertheless, the magnitude of $s_c$ decreases with the
stellar mass.  Also, $\tau_K$ is large at relatively large
$a_i$ such that the condition for adiabatically may be violated
during episodes of rapid mass loss when $\tau_M$ is reduced below
$\tau_K$. Nevertheless, planets and planetesimals can still be
retained at large distances provided the fractional loss of mass is
limited during these brief events. Substantial mass reduction of the
host stars can also cause dynamical instability of multiple-planet
systems (Debes \& Sigurdsson 2002). We will further explore this
possibility in a subsequent paper.

\subsection{Entrainment of small planetesimals by the gas outflow}

We demonstrate the orbital evolution of marginal and small
planetesimals with numerical solutions of the equation of motion. In
general, the orbits of intermediate ($s \sim 1$ km) size, more
distant ($r \sim 10$ AU) planetesimals may enlarge at rates slower
than the adiabatic expansion rates for the more massive and distant
planetesimals.

If a planetesimal is comparable to or smaller than a critical size,
orbital decay would be introduced due to gas drag. The gas drag
would reduce the planetesimal's momentum notably, and the angular
momentum would not remain constant. From equations (\ref{eq:motion})
and (\ref{eq:fdrag}), we find that drag by the outwardly flowing gas
exceeds the stellar gravity on particles smaller than
\begin{equation}
s_{\rm blowout} \simeq { r \rho_{wind} v_{wind}^2 \over \rho_p
v_K^2} \sim {{\dot M_\ast} P_K \over \rho_p r^2} {v_{wind} \over
v_K} \label{eq:sblowout}
\end{equation}
where $P_K$ is the orbital period. These very small particles ($s <
s_{\rm blowout} \sim 1$ cm) are entrained by the stellar wind.

\begin{figure}
  \includegraphics[angle=-90, width=150mm] {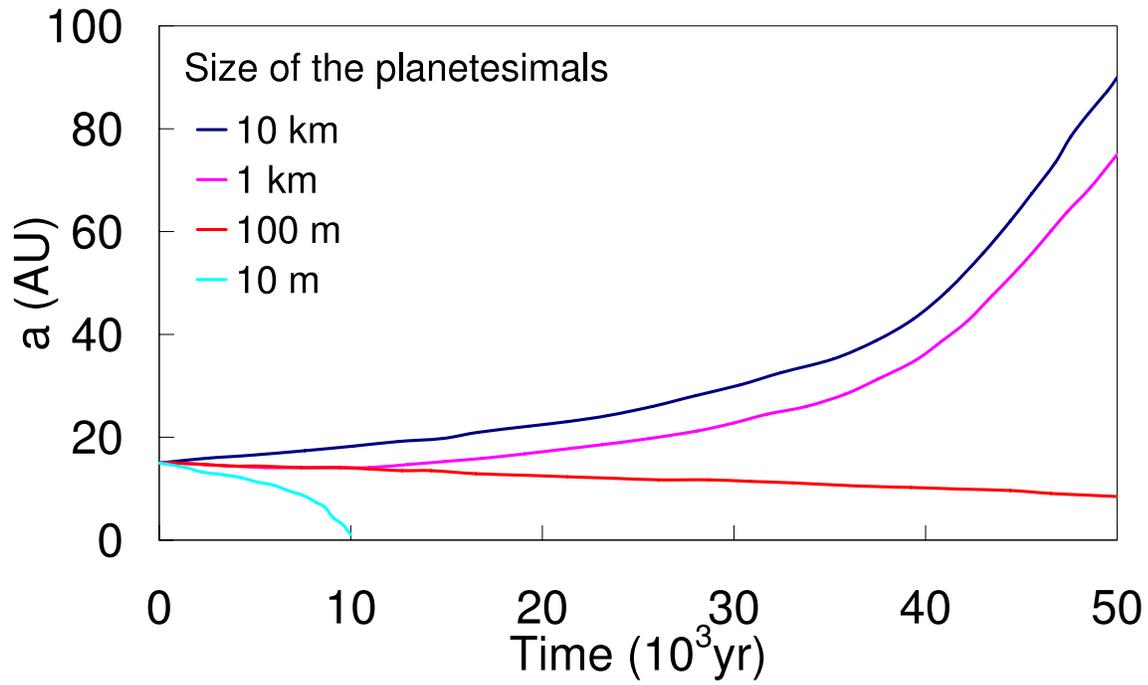}\\
  \caption{Orbital evolution of 4 planetesimals. The mass of the star
  changes linearly from an initial value $M_i = 3$ $M_\odot$ to a final value
  $M_f = 0.5$ $M_\odot$ in 50,000 years. Curves show the evolution of
  semi-major axis as a function of time for 4 objects all starting from 15 AU,
  with size $s=$ (from top to bottom) 10 km, 1 km, 100 m and 10 m. The wind velocity is $v_{wind} = 10$ km s$^{-1}$}
  \label{fi:divergent_evolution}
\end{figure}

In order to quantitatively illustrate the divergent evolutionary
paths of these particles, we consider a series of 4 planetesimals
with radius of 10m, 100m, 1km and 10km all starting at 15Au from the
star, whose orbits are initially without eccentricity. The mass of
the star changes linearly from an initial value $M_i = 3$ $M_\odot$
to a final value $M_f = 0.5$ $M_\odot$ in 50,000 years. The orbital
evolution of these planetesimals is shown in Figure
\ref{fi:divergent_evolution}. The 10 km object's has a final orbital
radius 6 time of the its initial value, which is exactly the
reduction factor of the stellar mass. This correlation implies that
the planetesimal's orbital evolution is adiabatic. The $s=1$ km
object also moves outward, but its expansion factor is considerably
smaller than that of the 10 km-size planetesimal. The 100m object
undergoes a small amount of orbital decay and its orbital evolution
is stalled at 10 AU. The 10 m planetesimal quickly spiral into the
central region and it is expected to be evaporated at about 10,000
years.

The processes of orbital decay (for planetesimals with $s_{\rm
blowout} < s < s_c$) and entrained outflow (for planetesimals with
$s < s_{\rm blowout}$, due to the hydrodynamic drag by the
outflowing gas, are equivalent to the Poynting-Robertson (PR) drag
and radiative blow out by stellar radiation. During the AGB mass
loss phases, hydrodynamic effects are much more intense than
radiative effects.

\subsection{Capture of small planetesimals by the planets' mean-motion
resonances}

In the above example, we neglected the presence of any major
planets. Such planets also undergo adiabatic expansion during the
planetary- nebula phase when their host stars lose a substantial
fraction of their masses. For these large entities, the orbital
expansion is adiabatic and take place on a time scale $\tau_{\rm
exp} \simeq \tau_M$. In addition, they exert dynamical perturbations
on the nearby planetesimals, especially those on their low-order
mean-motion resonances where their orbital periods are commensurate
with each other in terms of small integers.

During the stellar mass loss, planets' orbits expand from their
initial location $r_p$ to their final location $r_{\rm final} =
(M_i/M_f) r_p$. Beyond these planets' orbits, small or marginal-size
planetesimals, initially located at $a_i$ (we will stick to $r$ for
the radius of planet, since it always keeps a circular orbit, and
$a$ for the semi-major axis for planetesimals.), either undergo
orbital decay or expand at slower paces. Since their differential
motion is determined by the gas drag effect, a planet and a (more
distant) planetesimal cross each other's orbit on a time scale
\begin{equation}
\tau_\Delta \simeq {(a_i-r_p) \over a_i} \tau_{\rm drag} (s) = {(a_i
- r_p) \over a_i} {s \over s_c} \tau_M.
\end{equation}
Provided their initial location $a_i < r_{\rm final}$, particles
with sizes between
\begin{equation}
s_{\rm catchup} \sim { s_c a_i \over (a_i - r_p)}
\end{equation}
and $s_{\rm blowout}$ generally undergo orbit crossing with the planet
because the expansion of their orbits is slower than that of the planet.

Planetesimals along a planet's expansion path encounter its
mean-motion resonances prior to their orbit crossing. Provided the
duration $\tau_\Delta$ across the width of these resonance
($\Delta$) is long compared with the libration period near their
centers ($\tau_{\rm lib}$) planetesimals would be captured onto the
planets' mean-motion resonances (Murray \& Dermott 1999). For
example, a planet's 2:1 resonance has a finite effective width
$\Delta / r_p \simeq 2 {\rm exp 1/2} (M_p/M_\ast)^{1/2}$, where $P_K
= 2 \pi \tau_K$ is the orbital period of the planet, and a libration
period at its center $\tau_{\rm lib} \simeq (M_\ast / M_p) ^{1/2}
P_K/2 {\rm exp (1/2)}$. For this 2:1 resonance, the critical condition for
capturing a planetesimal (with a size $s$) is
\begin{equation}
M_p > {s_c \over s} {P_K \over 4 \tau_M {\rm exp 1}} M_\ast.
\label{eq:mcap}
\end{equation}
Planetesimals with $s_{\rm catchup} < s < s_{\rm rescap}$ are
captured onto the planet's 2:1 resonance where
\begin{equation}
s_{\rm rescap} \simeq s_c {P_k \over 4 \tau_M {\rm exp 1} } {a_i
\over 1.6 r_p} {M_\ast \over M_p}. \label{eq:rescap}
\end{equation}
In the above expression, the term $a_i/1.6 r_p$ is introduced to
take into account of the stellar mass loss which is the cause of
planet's orbital expansion. Equation (\ref{eq:mcap}) indicates that
prior to reaching $\sim 50-80$ AU, a Jupiter-mass planet would be
able to capture exterior km-size planetesimals ($s \simeq s_c (\sim$
1 km)) onto its 2:1 resonance. An Neptune-mass ice giant would also
be able to capture these marginal-size planetesimals while its
semi-major axis is $< 10$AU.

Planetesimals with initial semi-major axis $a_i < 1.6 r_p$ are
already interior to the planet's outer 2:1 resonance and they cannot
be captured onto it. Also, relatively small ($s < s_c$)
planetesimals are unlikely to be captured onto the planet's 2:1 mean
motion resonances because their orbital decay is too rapid. However,
this special location is the most distant lowest-order mean-motion
resonance from the planet. There are many other lowest-order (such
as 3:2 and 4:3) mean-motion resonances which are closer to the
planet's orbit. At these location, planet's resonant capture
probability is enhanced.

In the proximity of the planet, these lowest-order mean-motion
resonances overlap with each other and planetesimals with negligible
eccentricities and $\Delta \simeq (12)^{1/2} R_R$ can enter into the
planet's Roche radius ($R_R = (M_p/3 M_\ast)^{1/3} r_p$) within a
synodic period $\tau_{\rm syn} \simeq (3 r_p/ 2 \Delta) P_K$
(Lissauer 1987). Using a similar argument as above, the capture
criterion becomes $\tau_\Delta > \tau_{\rm syn}$, {\it i.e.} the
planetesimals must not be able to diffuse through this ``feeding
zone'' before they had an inferior conjunction. Thus, a fraction of
planetesimals with
\begin{equation}
s < s_{\rm bypass} = {s_c \over 18} \left( {3 M_\ast \over 2 M_p }
\right)^{2/3} {P_K {\dot M_\ast} \over M_\ast} \label{eq:sbypass}
\end{equation}
would be able to bypass the planet's expanding orbit. For Jupiter-mass
planets at $\sim 30 AU$ around typical intermediate-mass AGB stars,
$s_{\rm bypass} \sim (1-10) s_c$.

Although YORP effects can modify the radial distribution of km-size
planetesimals during the host stars' main-sequence phase (see
\S\ref{sec:model}), we neglect it in the determination of $s_{\rm
catchup}$, $s_{\rm rescap}$, and $s_{\rm bypass}$.  This
approximation is adequate for the AGB mass loss phase provided the
host stars' $\dot M > \dot M_Y$ (see eq \ref{eq:mdotyorp}).  During
the mass-loss process, a planet's orbit expands by a factor of
$M_i/M_f$. Planetesimals with sizes in the range
\begin{equation}
s_{\rm bypass} < s < s_{\rm catchup}
\end{equation}
are shepherded by the planet along its path. This size range is
approximately an order of magnitude larger and smaller than $s_c$. A
population of slightly larger ($s_{\rm rescap} < s < s_{\rm
catchup}$) and more distant (with initial $a_i > 1.6 r_p$)
planetesimals are captured onto the planet's 2:1 mean-motion
resonances. Smaller or closer planetesimals may be also captured
onto its other (such as 3:2 and 4:3) mean-motion resonances.

\subsection{Numerical calculation of resonant capture process}

\begin{figure}
  \includegraphics[angle=-90, width=150mm] {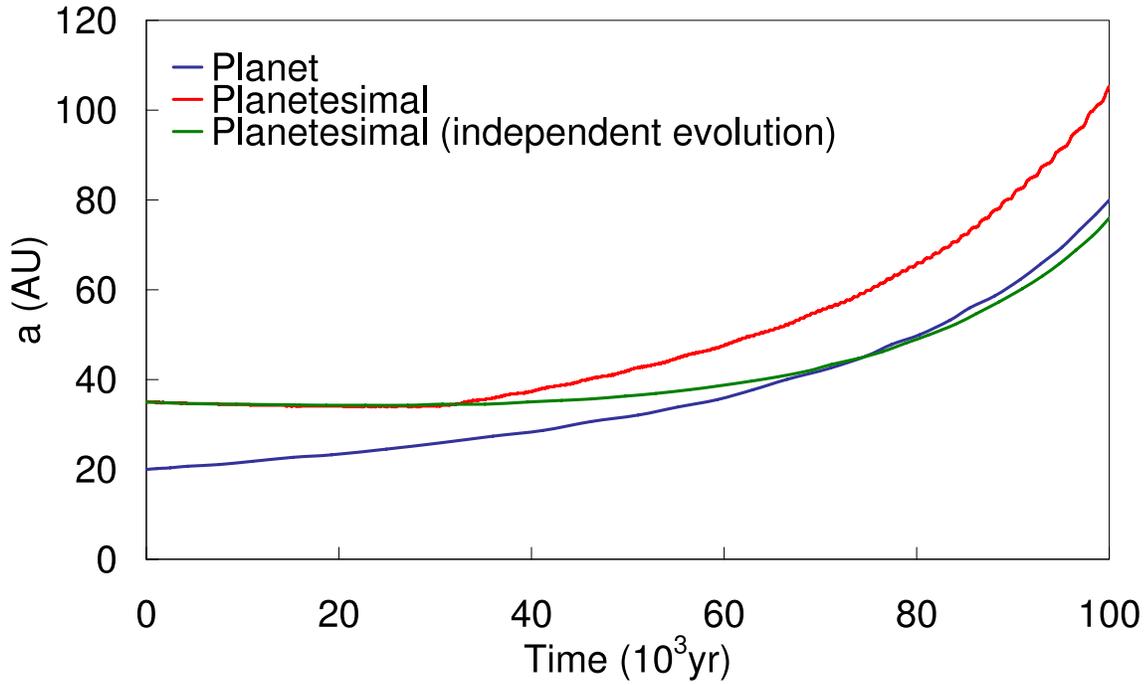}\\
  \caption{This plot compares the orbital evolution of a $s=$0.1 km planetesimal
  when there is a giant planet in the system (red) or not (green), while the blue curve
  shows the orbital evolution of the planet. The density of the
  planetesimal is $\rho_p = 3$ g cm$^{-3}$ and it starts out with a circular
  orbit at 35 AU around a $M_i =4$ $M_\odot$ star which linearly
  reduces its mass to $M_f = 1$ $M_\odot$ on a time scale of 100,000 yr
  through a wind with a speed $v_{wind} = 10$ km s$^{-1}$. The planet with a
  mass $M_p = 1$ $M_J$ was placed in a circular orbit at 20 AU initially.}
  \label{fi:capture_a}
\end{figure}

In order to illustrate the resonant capture process, we carried out
a series of numerical calculations. In Figure \ref{fi:capture_a}, we
plot the orbital evolution of a 0.1 km-size planetesimal. It has a
density $\rho_p = 3$ g cm$^{-3}$ and it starts out with a circular
orbit at 35 AU around a $ M_i =4 M_\odot$ star which linearly
reduces its mass to $M_f = 1 M_\odot$ on a time scale of 100,000 yr
through a wind with a speed $v_{wind} = 10$ km s$^{-1}$. A planet
with a mass $M_p = 1$ $M_J$ was placed in a circular orbit at 20 AU
initially. We also present the orbital evolution of the planetesimal
when there is no giant planet in the system for comparison.

\begin{figure}
  \includegraphics[angle=-90, width=150mm] {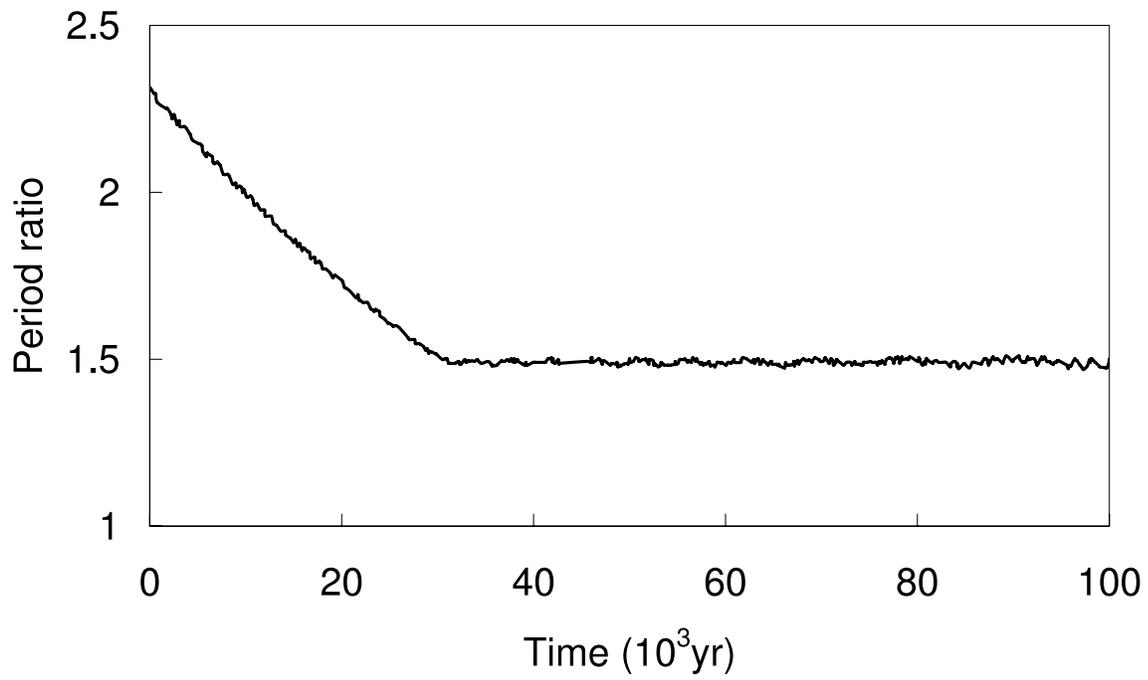}\\
  \caption{The evolution of period ratio ($P_{planetsimal} \over P_{planet}$)
  verse time for the system in Figure \ref{fi:capture_a}.}
  \label{fi:capture_p}
\end{figure}

As mentioned before, the orbital radius of the giant planet only
depends on the central stellar mass and initial orbit and
adiabatically evolve from 20 AU to 80 AU. The planetesimal's
semi-major axis first decreases a little bit until about 30,000
years, when it is captured by the 3:2 mean-motion resonance with the
giant planet. It's also the time that the evolution tracks of the
planetesimal with and without a giant planet in the system diverges.
Before the capture point, the two tracks almost exactly overlap each
other, while after that time, the resonance interaction pushes the
planetesimal to move outward. Meanwhile, the ratio of the
planetesimal's semi-major axis to the planet's becomes fixed on some
value (in this case, the ratio is $1.5^\frac{2}{3}$). Figure
\ref{fi:capture_p} shows the evolution of their period ratio.
Starting from about $2.3:1$ the ratio of the two periods decreases
until the resonance point 3:2, and then it remains fixed on this
value. During this process, although the period ratio passes 2:1 at
about $10,000$ years, the planetesimal was not captured by the
planet's 2:1 mean motion resonance.

\begin{figure}
  \includegraphics[angle=-90, width=150mm] {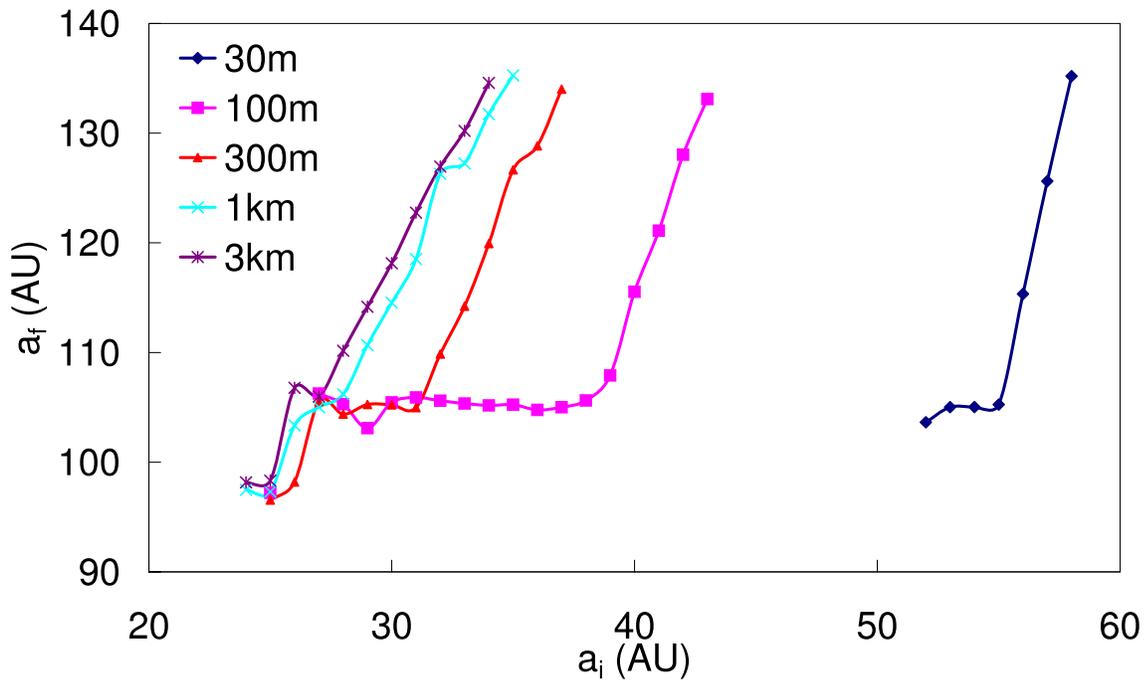}\\
  \caption{The orbital evolution of a group of different-size
  planetesimals verse time. The size of the planetesimals in the five
  curves are (from left to right) 3 km, 1 km, 300 m, 100 m and 30 m. The system
  is the one in Figure \ref{fi:capture_a}.}
  \label{fi:af_ai}
\end{figure}

As we have mentioned above that there are many factors which
determine planetesimal's resonant capture probability. First, for
planetesimals with certain size, the initially positions of the
planet and the planetesimals are important. Once the planet's
original orbit is set, only a range of initial position is
appropriate for the planetesimals which would be captured. In Figure
\ref{fi:af_ai}, we show the relation between initial positions and
the final positions (semi-major axis) during the orbital evolution
of several planetesimals with a range of sizes. In this model we
adopt the same planetary system as in Figure \ref{fi:capture_a}:
with a linear mass loss prescription for the host star (a constant
loss rate from $M_i = 4$ $M_\odot$ to $M_f = 1$ $M_\odot$ within
100,000 yr and $v_{wind} = 10$ km s$^{-1}$). A $1 M_j$ planet is
placed on an initial circular orbit with a radius $20$ AU, and its
orbit evolves to $r_f=80$ AU, according to the ratio of the star's
mass loss. All planetesimals have density $\rho_p =3$ g cm$^{-3}$.
This figure shows that planetesimals with size in the range of
3km-30m could be captured onto the planet's mean-motion resonance,
including 2:1, 3:2 and 4:3 resonant point (period ratio). It shows
that despite a wide range of initial positions, for planetesimals of
the same size, only the ones with initial $a_{i}$ in a "critical
range" could be captured onto the planet's mean motion resonances.
For example, among all the $s=300$ m planetesimals starting from 25
AU to 37 AU, only the ones with $a_i$ between 27 AU and 31 AU could
be captured onto 3:2 mean motion resonance.

Second, a planetesimal's size is also a determining factor for
resonant capture probability. Equation (\ref{eq:rescap}) indicates
that the effective size range for resonant capture ($s_{\rm rescap}
< s < s_{\rm catchup}$) depends on the planetesimal's initial
location $a_i$, so the "critical range" of initial positions in the
first point depends on the size of planetesimals. This dependence is
confirmed by our numerical calculations. In Figure \ref{fi:af_ai}
the "critical range" for $s=1$ km object to be captured onto 3:2
resonance is from 26 AU to 28 AU, while for $s=100$ m object is from
26 AU to 38 AU. The smaller planetesimals comparatively feel
"stronger" gas drag than larger ones at the same orbit, so they must
have larger initial $a_i$'s so that the gas drag is insufficient to
drag them in and they have the possibility to being captured by the
planet. Generally speaking, this "critical range" is larger and
further for small particles, as the picture demonstrates. These
results are in good agreement with the above analytic estimates.

The magnitude of the gas drag is proportional to $\rho_{wind}$ which
is determined by the velocity of the gas wind and the mass loss rate
of the central star. However, equation (\ref{eq:sc}) indicate that
$s_c$ is essentially independent of $\dot M_\ast$ and $v_{wind}$.
Nevertheless $s_{\rm rescap}$ does depend on $\tau_M$ and the star
to planet's mass ratio as well as planetesimals' internal density
(through $s_c$). In order to explore these dependence, we consider
variation in the model parameters from the standard model we have
been working with above, {\it i.e.} with $\dot M \sim 10^{-4}
M_\odot$ yr$^{-1}$ and $v_{wind} = 10$ km s$^{-1}$. These quantities
may vary not only from star to star but also temporally around any
host star. In addition, the initial and final mass of the stars
determines the extent of planet's orbital expansion and sweep up of
the residual planetesimals. The internal density of the
planetesimals ($\rho_p$) may also have a range of possible values,
from $\sim 3$gm cm$^{-3}$ for asteroids to an order of magnitude
smaller for comets.

Apart from the model parameters we have discussed above, we have
adopted several different prescription for the evolution of the
mass loss rate including constant, exponentially increasing, or
ramp ups followed by declines.  We find that once we have specified
the initial and final stellar mass and the duration of the mass loss,
our results are insensitive to variations in the mass loss rates.
We also consider a range of planetary masses ($1-10 M_j$) and
eccentricities (0 and 0.1).  Our results are also insensitive to
the magnitude of these parameters.

In the present context, a planet's final semi-major axis is
generally less than 100 AU and its asymptotic period is $< 1000$ yr.
Thus, their orbital expansion is always adiabatic such that their
orbits expand by a factor of $M_i/M_f$. By specifying the planet's
initial orbital radius, we can assess the inventory of planetesimals
to be captured by the planet.

At which ratio of the orbital circle of the planetesimal and the
planet would the particles be fixed is also determined by many
factors. The rate of the orbital circle of the planetesimal and the
planet would keep rising while no resonance happens. When it close
to a resonance point, whether or not it could stop raising and stay
is primarily determined by whether or not the resonance is strong
enough, which is decided by factors like the gas' density, the
positions of the planetesimal and the planet, etc. In our
discussion, there are three major resonance points: rate of the
orbital circle equals to 3:2, 4:3 and 3:2, while 3:2 is the most
important one. It's shown in Figure \ref{fi:af_ai}. Take the curve
of $s=300$m planetesimals for example: the objects starting from
$a_i=$27 AU-31 AU are captured onto 3:2 resonance, with $a_f$ around
105 AU; starting from $a_i=$25 AU-26 AU are captured onto 4:3
resonance, which have $a_f$ around 97 AU; and starting from $a_i=$35
AU-36 AU are captured onto 2:1 resonance, which have $a_f$ around
128 AU.

\section{Collisional evolution of planetesimals and production of grains}

In the previous section, we suggest that km-size icy particles are
the preferred population of residual planetesimals for planets'
resonance capture during the AGB mass loss phase of their host
stars. In the next section, we first estimate the condition under
which a population of such planetesimals may emerge at 10-100 AU
prior to the AGB phase.

We then show that as a consequence of their orbital expansion and
planets' sweeping process, these planetesimals' eccentricities
increases to modest values and their collision frequency is strongly
enhanced. We show that planetesimals orbit crossing with each other
leads to frequent high-velocity disruptive impacts and the
production of small grains. These grains can then reprocess the
stellar radiation and produce the observed FIR rings around the
white dwarf at the center of the helix nebula.

\subsection{Inventory of planetesimals}

In \S2, we indicated that around typical progenitor of planetary
nebula (A main sequence star), there may be a rich population of
planetesimals. We now estimate their characteristic size prior to
the mass loss from their host star.

There are two possible formation epochs for residual planetesimals
(first few Myr and through out the main sequence life time of their
host stars $\tau_\ast$). First-generation planetesimals are formed
in gas-rich protoplanetary disks. In such a gaseous environment, the
velocity dispersion of the small planetesimals ($\sigma$) is excited
by their mutual interaction and damped by their interaction with the
disk gas (Kokubo \& Ida 2002). Due to hydrodynamic drag by the
turbulent disk gas, these particles also undergo orbital decay
(Supulver \& Lin 2000). During this formative stage for their host
star, largest planetesimals can grow to sizes well beyond a few km,
especially in special locations such as the snow line and in the
proximity of proto gas giants. They can also be scattered to large
distances by the emerging planets (Zhou \& Lin 2007). Nevertheless,
we have already showed in the last section that during the planetary
nebula stage, planetesimals with $s > s_{\rm catchup}$ retain their
semi-major axis ratio with respect to any embedded planet. They are
unlikely to be captured onto any planet's expanding mean-motion
resonances.

Planetesimals can also grow {\it in situ} during the main sequence
life span of their host stars ($\tau_\ast \sim 10^{8-9}$ yr), long
after the depletion of the disk gas. In this case, their $\sigma$ is
determined by their own collisional properties. Collisions can lead
to both mergers and fragmentation. The condition for collisional
fragmentation for basalt and ices have been extensively studied
experimentally and simulated numerically (Stewart \& Leinhardt
2008). In collisions between two comparable km-size aggregates, the
critical impact speed $v_{\rm frag}$ for significant fragmentation
is around a few meter per second (which is comparable to their
surface escape speed $v_g$ for these planetesimals). Shattering
collisions are not only inhibits the coagulation of planetesimals
but also introduces considerable energy dissipation. In Keplerian
disks (such as Saturn's rings), particles' velocity dispersion
($\sigma$) is determined by its excitation due to elastic scattering
and damping due to inelastic collisions (Goldreich \& Tremaine 1977,
Bridges {\it et al} 1984). In gas-free disks with $\sigma < v_{\rm
frag}$, planetesimals undergo cohesive collisions with growing
$\sigma$ (Palmer {\it et al.} 1993, Aarseth {\it et al.} 1993). When
their $\sigma$ exceeds $v_{\rm frag}$, collisions quench both mass
and $\sigma$ growth. In the absence of a dominant population of
large planetesimals or embedded planets, planetesimals establish a
collisional equilibrium in which their $\sigma$ is approximately
isotropic with $\sigma \sim v_{\rm frag}$. An isotropic velocity
dispersion implies that planetesimals can only collide with others
in an annulus with a half width $a_i \sigma /v_K$ and the magnitude
of $\sigma$ determines not only the collisional frequency but also
the thickness of the planetesimal disk.

For km-size planetesimals, $v_{\rm frag} \sim v_g$ such that the
collisional cross section of these planetesimals is mostly
determined by their physical size. If most collisions are cohesive,
the growth time scale ($ \tau_{\rm growth}$) would be comparable to
the collisional time scale ($\tau_{\rm col}$)
\begin{equation}
\tau_{\rm growth} \sim \tau_{\rm col} \simeq (a_i/s)^2 (M_e / M_{\rm
tot}) (P_K/ 2 \pi) \label{eq:taugrow}
\end{equation}
where $M_e$ and $M_{\rm tot}$ are the individual and characteristic
total (at $a_i$) masses of planetesimals. The present estimates for
$M_{\rm tot}$ in the Kuiper Belt ranges from a fraction to a few
$M_\oplus$ and that of debris disks around A stars may be near the
upper end of this range (see \S2). From equation (\ref{eq:taugrow}),
we find the characteristic size of planetesimals to be
\begin{equation}
s_{\rm char} \simeq \left( 30\ {\rm AU} \over a_i \right)^{1/2}
\left( 1\ {\rm g \ cm}^{-3} \over \rho_p \right) \left( \tau_{\rm
growth} \over \tau_\ast \right) \left( \tau_\ast \over 0.1\ {\rm
Gyr} \right) \left( M_{\rm tot} \over 1\ M_\oplus \right) km.
\label{eq:schar}
\end{equation}
The above equation suggests that the spontaneous production of an
isolated population of km-size planetesimals within the main sequence
life span of a A star requires $M_{\rm tot}$ to be a few $M_\oplus$.
This estimate is consistent with that simulated for the Kuiper Belt
region (Kenyon \& Bromley 2004).

In the derivation of $s_{\rm char}$, we assume that $\tau_{\rm
growth} \sim \tau_{\rm col}$, {\it i.e.} all collisions are
cohesive. Since $\sigma \sim v_{\rm frag}$ some collisions are
disruptive and $s_{\rm char}$ represents an upper limit in the
planetesimals' size distribution. In \S2, we briefly show that the
particle size distribution may follow a power law in which
\begin{equation}
dN/ds = (N_o/s_{\rm max}) (s/s_{\rm max})^{-3.5}
\label{eq:mrn}
\end{equation}
where $s_{\rm max}$ is the size of the largest particles and $N_o$
is a normalization factor. With this size distribution, most of the
mass are attained by the largest planetesimals. As long as $s_{\rm
char} > 1$ km, there is always a supply of residual planetesimals
which may be captured by the resonance of expanding planets.

Equation (\ref{eq:schar}) implies a large potential dispersion in
$s_{\rm char}$ around stars with different $\tau_\ast$ and $a_i$.
Perhaps the largest uncertainty is the total mass of residual heavy
element $M_{\rm tot}$. Around stars with relatively massive debris
disks, $s_{\rm char} > 1$ km and the total mass of km-size
planetesimals is
\begin{equation}
M_{\rm km} \sim (1 {\rm km}/ s_{\rm char})^{1/2} M_{\rm tot}
\propto M_{\rm tot} ^{1/2}.
\end{equation}
But, in less massive disks, the supply of km-size planetesimals may
be limited. Based on this consideration and the uncertain fraction
of A stars with planets, we do not expect the detection of trapped
planetesimals around the remnant white dwarf at the center of every
planetary nebula.

\subsection{Eccentricity excitation}

After km-size planetesimals are captured onto the planet's
mean-motion resonances, their orbital expansion would be locked
provided the orbital migration time scale is longer than the
resonant libration time scale {\it i.e.} $\tau_M > \tau_{\rm lib}$
or equivalently,
\begin{equation}
M_p > (M_\ast / 4 {\rm exp} 1) (P_K/\tau_M)^2 \sim 10^{-5} M_\ast.
\end{equation}
Similar processes are found in the tidal evolution of Galilean
satellites (Goldreich 1965, Lin \& Papaloizou 1979, Peale \& Yoder
1981) and migration of resonant planets (Lee \& Peale 2002, Kley
 {\it et al.} 2004).

During their forced co-moving orbital expansion, the conservation of
an adiabatic invariance induces the growth of resonant planetesimals'
eccentricity ($e_s$) (Murray \& Dermott 1999) on a time scale
\begin{equation}
\tau_{e, e} = e_s /{\dot e}_s = 2 e_s ^2 \tau_{\rm exp} = 2 e_s^2
\tau_M.
\end{equation}
The above expression is for planetesimals in the planet's 2:1 mean
motion resonance. A similar expression can be derived for planet's
other mean-motion resonances.

Although the stellar outflow is in the radial direction, its drag on
the planetesimals can also lead to the damping of their eccentricities
on a time scale
\begin{equation}
\tau_{e, d} \simeq (s / s_c) \tau_M.
\end{equation}
The above expression is derived from a linear perturbation analysis
of equation (\ref{eq:fdrag}) in the limit of small $e_s$. Comparison
between $\tau_{e, e}$ and $\tau_{e, a}$, it is apparent that
planetesimals' $e_s$ grows initially until an asymptotic equilibrium
is established in which
\begin{equation}
e_e \sim (s/ s_c)^{1/2}.
\end{equation}
After they are captured by the planet at a location $r_p = r_{\rm
cap}$, the planetesimals' eccentricity growth may also be quenched by
the termination of the stellar mass loss with an asymptotic
\begin{equation}
e_a \simeq ({\rm ln} (r_{\rm final} / r_{\rm cap}) )^{1/2}.
\end{equation}

In general, the eccentricity of co-moving resonant planetesimals is
excited to the minimum of $e_e$ and $e_a$ which may both be a
significant fraction of unity. (Both of these estimates on the
asymptotic eccentricities break down as their values approach to
unity.) Since the semi-major axes of the resonant planets are in
lock-steps with that of the planet, the orbits of widely separated
planetesimals overlap. The onset of orbit crossing creates
opportunities for collisions between these planetesimals.

The eccentricities of distant planetesimals with period $P_K \sim
\tau_M$ (at $a_i \geq 500$ AU) may also increase as a consequence of
their marginally non-adiabatic adjustment to the host star's
declining gravitational potential. However, these planetesimals'
peri-center distances would not decrease significantly so that they
are unlikely to venture into the regions of interest. At these large
distances, planetesimals are sparsely populated, their collision
frequency is also expected to be low.

\begin{figure}
  \includegraphics[angle=-90, width=150mm] {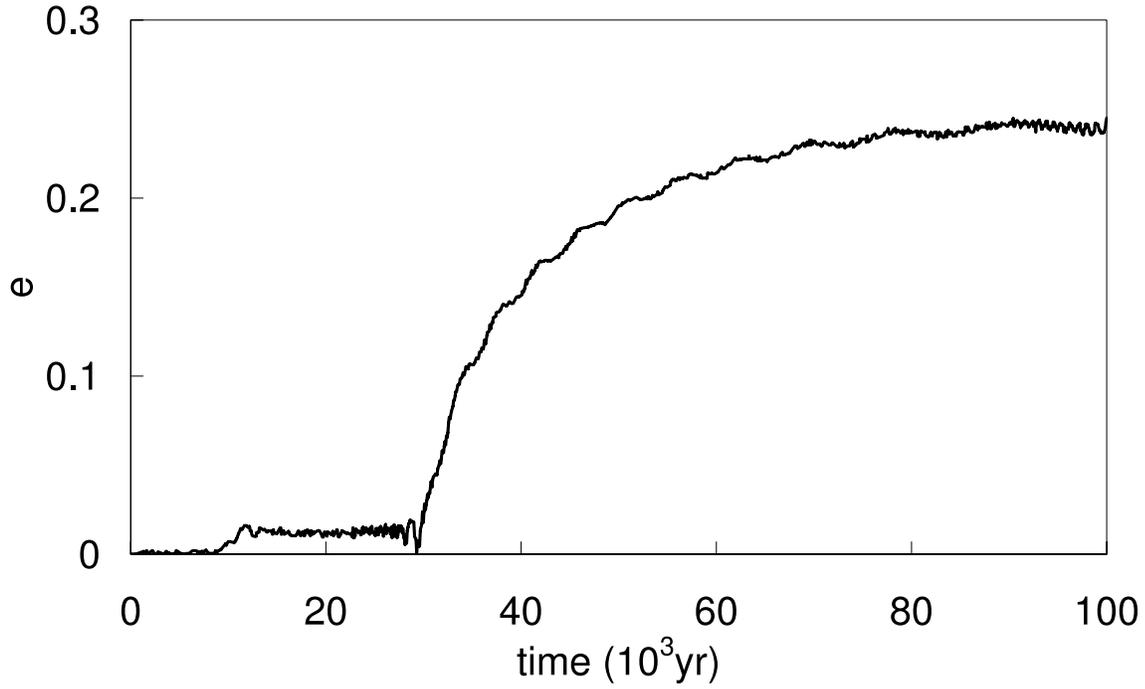}\\
  \caption{The eccentricity evolution of the 100m-size planetesimal in the system of
  Figure \ref{fi:capture_a}}
  \label{fi:capture_e}
\end{figure}

In order to verify these analytic estimates, we show in Figure
\ref{fi:capture_e} the evolution of the planetesimal's eccentricity
in the system in Figure \ref{fi:capture_a}. In this case, it remains
negligible until $10,000$ yr. when it encounters the planet's 2:1
mean-motion resonant point. Although not being captured, its
eccentricity still slightly increases. When the planetesimal is
captured onto the planet's 3:2 mean-motion resonance at $30,000$
years, its eccentricity is evidently excited. At the end of our
simulation it reaches its asymptotic value $\sim 0.24$. These result
is in good agreement with the above discussions.

\subsection{Collisional frequency and outcome}

\begin{figure}
  \includegraphics[angle=-90, width=150mm] {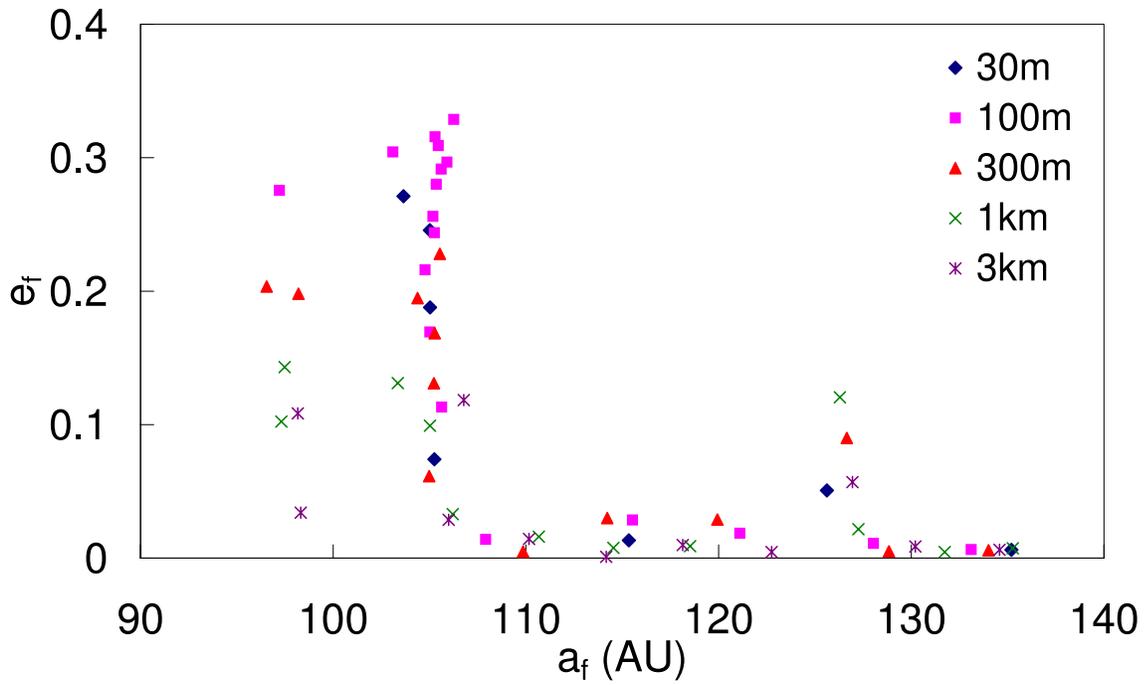}\\
  \caption{The final eccentricity verse final semi-major
  axis for several planetesimals with different sizes in the system
  of Figure \ref{fi:af_ai}.}
  \label{fi:ef_af}
\end{figure}

The results in Figure \ref{fi:capture_e} can be generalized. In
Figure \ref{fi:ef_af}, we show the asymptotic eccentricity
distribution of a population of planetesimals (the system in Figure
\ref{fi:capture_a}) verse their final orbital semi-major axis. It's
very clear, as we pointed out before, there are three major
resonance points of this system which can capture planetesimals onto
co-moving orbits with giant planet and excite their eccentricities:
3:2 resonance point (the most important one, final semi-major axis
of planetesimals $a_f \sim 105$ AU); the 2:1 resonance point ($a_f
\sim 97$ AU) and 4:3 resonance point ($a_f \sim 128$ AU). These
three resonance points correspond to the three vertical bands on the
plot. Once being excited, the eccentricity of the planetesimals will
increase to about $0.1 \sim 0.3$. For planetesimals which are
captured onto 3:2 resonant orbit, since their asymptotic semi-major
axis is around 100 AU their orbits cross each other in the region
between $\sim 80-120$ AU. Thus the planetesimals' eccentricity
excitation enlarges the radial extent of their impact zone from
$\sim a_i v_{\rm frag}/v_K \sim 10^{-3} a_i$ to $\sim 2e_s a_i \sim
0.4 a_i$. This expansion greatly increases the pool of potential
impactors.

As planetesimals along the expanding paths of the planet are
captured onto its mean-motion resonances and co-moved by it to large
distances, the surface number density $\Sigma_p$ of planetesimals
can increase by a factor $f_\Sigma$ ($\sim$ a few), depending on the
initial distribution of $\Sigma_p$ prior to the star's mass loss.
(In this estimate, we took into account the growth of planetesimals'
eccentricity associated with their resonantly forced orbital
expansion.)

In contrast, the planetesimals' dispersion velocity in the direction
normal to the disk plane does not change significantly (from its
initial value $\sigma \sim v_{\rm frag}$) because the forcing by the
mean-motion resonances is applied in the radial direction. (The
velocity dispersion is no longer isotropic, {\it i.e.} the radial
component of the velocity dispersion ellipsoid, $\sigma_r \sim e_s
v_K$, becomes much larger than that in the direction normal to the
orbital plane, $\sigma_z \sim v_{\rm frag}$.) Consequently, the
spacial density of resonant planetesimals ($n_p^\prime$) actually
increases from their initial values ($M_{\rm tot}/ (M_e 2 \pi a_i^2
H)$ where $H \sim \sigma_z a_i / v_K$ is the disk thickness) by a
factor similar to $f_\Sigma$ and the collisional frequency is
enhanced by both the density and velocity enhancement. The
characteristic collisional time scale $\tau_{\rm col}$ for the
km-size planetesimals (which are trapped in the planet's mean-motion
resonances) is reduced from its original magnitude to
\begin{equation}
\tau_{\rm col} ^\prime \sim 1/(n_p^\prime \sigma_r s^2) \sim
\tau_{\rm col} (v_{\rm frag} /e_s v_K f_\Sigma) (1 \ {\rm km} / s_c)
^{1/2} \sim (10^{-3}-10^{-4}) (1 \ {\rm km} / s_c) ^{1/2} \tau_\ast.
\end{equation}
In the above expression, we assume that within the main sequence
life span ($\tau_\ast$) of the host star,{\it i.e.} prior to the AGB
mass loss phase, perfectly cohesive collisions led to the formation
of planetesimals with the power-law size distribution in equation
(\ref{eq:mrn}) up to a size $s_c$.

During and after the AGB mass loss phase, the radial velocity
dispersion associated with the asymptotic mean eccentricity ($\sim
0.2$) is $\sim 0.5$km s$^{-1}$ at $\sim 100$ AU. Since this value of
$\sigma$ is much larger than $v_{\rm frag}$, we anticipate all
collisions during this stage will lead to powderization of the
planetesimals. Since only km-size planetesimals are resonantly
captured and co-move with the orbital expansion of the planet,
production rate of fragment would be
\begin{equation}
{\dot M}_{\rm frag} \sim (1 \ {\rm km} / s_c) ^{1/2} M_{\rm tot} /
\tau_{\rm col}^\prime \sim (e_s v_K f_\Sigma / v_{\rm frag} )
(M_{\rm tot} / \tau_\ast) \sim 10^{3-4} (M_{\rm tot} / \tau_\ast).
\end{equation}

With this erasion rate, the total mass of fragments produced during
the AGB mass loss phase would be
\begin{equation}
M_{\rm tot}^\prime \sim {\dot M}_{\rm frag} \tau_M \sim M_{\rm tot}
(10^{3-4} \tau_M / \tau_\ast). \label{eq:mtotprime}
\end{equation}
During the mass loss epoch, the magnitude of $\tau_{\rm col}^\prime$
can be reduced to $ \sim 10^{4-5}$ yr which would be comparable to
$\tau_M$. In these systems, a significant fraction of $M_{\rm tot}$
could be fragmented through these disruptive processes. Since the
emergence of km-size planetesimals requires $M_{\rm tot}$ to be a
few $M_\oplus$, we anticipate that in planetary nebula where they
coexist with some gas giants, the total mass of their fragmentary
debris ($M_{\rm tot}^\prime$) may indeed exceed that of the Earth.

Fragments' mass spectrum, immediately after the disruptive impacts,
depends on the highly uncertain material strength of planetesimals.
If these planetesimals are loosely bound aggregates high speed
($\sim$ km s$^{-1}$) impacts may disperse them down to the sizes of
their basic building blocks. For simplicity, we assume a power-law
size distribution as expressed in equation (\ref{eq:mrn}). In this
case, a major portion of $M_{\rm tot}^\prime$ is contained in the
largest fragments (with $s = s_{\rm max}$ and mass $m_{\rm max} = 4
\pi \rho_p s_{\rm max}^3 /3$) so that the normalization factor can
be derived from equation (\ref{eq:mrn}) to be $N_o = M_{\rm
tot}^\prime/ 2 m_{\rm max}$. In the absence of reliable experimental
data, we infer below the magnitude of $s_{\rm max}$ using the flux
of reprocessed radiation from the ring around WD 2226-210.

\subsection{Reprocessed radiation by the residual fragments}

Another implication of the power-law size distribution is the
fraction of stellar photons which may be absorbed and scattered by
the fragments. In the calculation on the intensity of reprocessed
radiation at any instance of time, we need to take into
consideration the potential depletion of all particles. We have
already indicated above that, during the gas loss phase, fragments
with sizes $s < s_{\rm blowout}$ are continually entrained and
removed by the outflowing gas from the location of their production
on a time scale $\tau_{\rm fly} \sim a_i/v_{wind}$. Similarly,
inward orbital decay of fragments in the size range $s_{\rm blowout}
< s < s_{\rm bypass}$ proceeds on a time scale $\tau_{\rm drag} =
(s/ s_c) \tau_M$ (equation (\ref{eq:taudrag2})) are not inhibited by
the planet's tidal barrier during their orbital decay.

In the calculation of intensity of reprocessed stellar irradiation
at a particular wavelength range ($\lambda$), we also need to
include a reduction factor ($s/\lambda$) for emissivity of small
particles (in the range from $s_\lambda \equiv \lambda > s$ down
to a minimum grain size $s_{\rm min}$). In the limit $s_{\rm bypass}
\sim s_c > s_{\rm blowout} > s_\lambda > s_{\rm min}$, the total area
covered by the fragments in the impact zone would be
\begin{equation}
\sigma_{\rm zone} = \left( \int_{s_{\rm bypass}} ^{s_{\rm max}} +
\int_{s_{\rm blowout}} ^{s_{\rm bypass}} {\tau_{\rm drag} \over
\tau_M} + \int_{s_\lambda} ^{s_{\rm blowout}} {\tau_{\rm blowout}
\over \tau_M} + \int_{s_{\rm min}} ^{s_\lambda} {s \over \lambda}
\right) \pi s^2 {d N \over d s} d s .
\end{equation}
Neglecting the YORP effect and inserting expressions for $d N/ d s$
in equation (\ref{eq:mrn}), we find that
\begin{eqnarray}
\sigma_{\rm zone} & \sim & {2 \pi N_o s_{\rm max} ^2} \bigg\{ \left[
\left(s_{\rm max} \over s_{\rm bypass}\right)^{1/2} -1 \right] +
{s_{\rm max} \over s_c} \left[ \left(s_{\rm bypass} \over s_{\rm
max} \right)^{1/2} - \left( s_{\rm blowout} \over s_{\rm max}
\right)^{1/2} \right] \nonumber \\
&  & +  {\tau_{\rm fly} \over \tau_M} \left[ 2 \left( s_{\rm max}
\over s_\lambda \right)^{1/2} - \left( s_{\rm max} \over s_{\rm
blowout} \right)^{1/2}  -  \left( s_{\rm max} \over s_\lambda
\right) \left( s_{\rm min} \over s_{\rm max} \right)^{1/2} \right]
\bigg\}. \label{eq:taugeneral}
\end{eqnarray}
Although AGB stars' $L_\ast$ can reach $\sim 10^{-3} L_\odot$ to
exert non-negligible YORP effect on km-size planetesimals at $\sim
10-100$ AU, this phase is brief and does not significantly affect
their orbits.

In the limit $s_c \sim s_{\rm bypass}$, $\sigma_{\rm zone} \sim 4
\pi N_o s_{\rm max}^{5/2} s_{\rm bypass} ^{-1/2}$.  Interior to the
impact zone and the planet's orbit, fragments with $s_{\rm bypass} >
s_{\rm blowout} > s_\lambda $ contributes to $\sigma_{\rm interior}
\sim (2 \pi N_o^\prime s_{\rm max}^3/s_c) [ ( s_{\rm bypass} /
s_{\rm max})^{1/2} - (s_{\rm blowout}/ s_{\rm max}) ^{1/2} ] \sim
2\pi N_o^\prime s_{\rm max}^{5/2} s_{\rm bypass} ^{-1/2} <
\sigma_{\rm zone}$ because the normalization factor $N_o ^\prime$
(which is determined by the resonant capture efficiency) is expected
to be smaller than $N_o$.  Exterior to the impact zone, small
fragments ($s < s_{\rm blowout}$) contributes to $\sigma_{\rm
exterior} \sim (2 \pi N_o s_{\rm max} ^2 \tau_{\rm fly} / \tau_M) [
2 ( s_{\rm max} / s_\lambda )^{1/2} - ( s_{\rm max} / s_{\rm
blowout} )^{1/2} - (s_{\rm max}/s_\lambda) (s_{\rm min} \ s_{\rm
max}) ^{1/2}] \sim \sigma_{\rm zone} (s_{\rm bypass}/s_\lambda)
(\tau_{\rm fly} / \tau_M)$ which may also be much smaller than
$\sigma_{\rm zone}$.

Equations (\ref{eq:sblowout}) and (\ref{eq:sbypass}) indicate that
both $s_{\rm blowout}$ and $s_{\rm bypass}$ increases with ${\dot
M}_\ast$. After the host star attains its $M_f$ (as in the case of
the white dwarf WD 2226-210 at the center of the Helix nebula), the
mass flux of gas outflow is significantly reduced. While the
eccentric resonant-captured planetesimals continue to collide with
each other on a time scale $\tau_{\rm col} ^\prime$ and generate
fragments at a rate ${\dot M}_{\rm frag}$. When both $s_{\rm
blowout}$ and $s_{\rm bypass}$ decreases below $s_\lambda$, equation
(\ref{eq:taugeneral}) is much simplified by a continuous power-law
size distribution (as in equation \ref{eq:mrn}) to
\begin{equation}
\sigma_{\rm zone} \sim {2 \pi N_o s_{\rm max} ^2} \left[
\left(s_{\rm max} \over s_\lambda\right)^{1/2} -1 \right].
\label{eq:taufull}
\end{equation}
In this case, the fragments are confined to the impact zone.

Finally, the particle size distribution may also be truncated at the
lower ranges the radiative blow out process ($s_\gamma$). In that
limit, $s_\lambda$ would be replaced by $s_\gamma$ in the above
expressions.

\section{Structure of the disk around WD2226-210}

The proximity of WD2226-210 at the center of the Helix nebula has
been imaged in multiple wavelength. The observed spectral energy
distribution (SED) has been analyzed in terms of a disk of dust
grains which are assumed to have the same astronomical size
distribution as equation (\ref{eq:mrn}) and a size range between
60-1000 $\mu$m (Su {\it et al.} 2007). (The lower size limit is set
($s_{\rm min} = 60 \mu m$) by the condition for radiative blow out
({\it i.e.} $s_\gamma$), albeit $s_{\rm max}$ is arbitrarily set.)
The total area covered by these dust grains is inferred to be
$\sigma _{\rm tot} \sim 6 \times 10^{27}$ cm$^2$ distributed over a
distance between 35-150 AU. For the assumed range of grain sizes,
their total mass is inferred to be $M_{\rm grain} \sim 0.1
M_\oplus$. (This mass is inferred from the assumed size distribution
in equation \ref{eq:mrn}, neglecting the contribution of
hydrodynamic blow out and drag on the small grains in equation
\ref{eq:taugeneral}). The magnitude of
\begin{equation}
M_{\rm grain} \sim 0.1 (s_{\rm max} / 0.1 {\rm cm})^{1/2} M_\oplus
\label{eq:mtotgrain}
\end{equation}
if the assumed size distribution extends to grains larger than 1000
$\mu$m=0.1 cm.) Since the present and past PR drag time scale for
these grains ($\sim 10$ My) is $< \tau_\ast$, they must be
continually replenished, presumably through the powderization of
their parent-body residual planetesimals. Su {\it et al.} (2007)
estimated the current collision time scale for $100 \mu$m grains to
be $\sim 0.1-1$ My which is $> \tau_M$.

Although this model provides a reasonable fit to the observed SED, a
challenging theoretical issue is whether the grains' parent bodies
may have sufficient collision frequencies to generate so many
fragments despite their (parent bodies') much smaller surface area
to volume ratio. If the generation of $M_{\rm grain}$ requires a
substantially more massive ($> > M_\oplus$) population of parent
bodies, its protracted retention during the main sequence life span
of their host stars would need to be addressed rather than assumed.

We now apply the results in the previous sections to provide some
quantitative justification for the model proposed by Su {\it et al.}
(2007). In \S2, we cite evidences that both gas giant planets and
debris disks are commonly found around A stars. In our model, we
consider the possibility that the progenitor star of WD 2226-210 was
surrounded by one or more gas giant planets (at distances comparable
to those around Uranus and Neptune) and a debris disk similar to the
Kuiper Belt in the solar system. In previous chapters, we have already
showed that in debris disks which contain a few (or more) $M_\oplus$
within $\sim 100$ AU, a population of km-size planetesimals (with a
total mass $M_{rm tot} > 1$ $M_\oplus$) can form through inelastic and
cohesive collisions (with a velocity dispersion $\sigma \sim v_{\rm
frag} \sim $ a few m s$^{-1}$) during the main sequence life span of
type A stars ({\it i.e.} within $\tau_\ast \sim 100$ My). Due to the
combined planetary tidal perturbation and the YORP effect, a fraction
of these residual planetesimals may accumulate near the planets' outer
mean motion resonances.

During the AGB mass loss phase when most of the stellar mass is
rapidly lost, semi-major axes of both planets and relatively large
residual planetesimals increases as $M_\ast^{-1}$. In contrast, the
outflowing wind imposes hydrodynamic drag on the planetesimals and
causes those with relatively small sizes ($s < s_c \sim 1$km) to
undergo orbital decay. When they cross the planet's expanding path,
these small planetesimals are captured onto its mean motion
resonances of the planets.

During the subsequent shepherd orbital expansion, the resonant
planetesimals are swept up with enhance surface densities (by a
factor $f_\Sigma \sim$ a few), excited eccentricities ($e_s \sim
e_e$ or $e_d$ which can be up to a significant fraction of unity),
and highly anisotropic velocity dispersion ($\sigma_r \sim e_s v_K
\sim 10^{2-3} \sigma_z$). The collisional time scale for these
planetesimals are greatly reduced (by a factor $\sim 10^{3-4}$) from
$\tau_\ast$ to $\sim \tau_M$ so that a significant fraction of the
planetesimals captured on the planet's mean-motion resonances may
actually collide with each other. The impact speed of these
collisions ($\sim \sigma_r \sim 0.5$km s$^{-1}$) is sufficiently
large to pulverize km-size planetesimals to very small pieces.

The actual size of the largest fragments ($s_{\rm max}$) is poorly
determined. However, we can infer it by assuming $M_{\rm grain}
\simeq M_{\rm tot}^\prime$ (from equations \ref{eq:mtotgrain} and
\ref{eq:mtotprime}) and $\sigma_{\rm tot} \sim \sigma_{\rm zone}$
(from equations \ref{eq:taugeneral} or \ref{eq:taufull}). If we
neglect any residual stellar winds and self-consistently adopt
$M_{\rm tot}^\prime \sim M_{\rm tot} = 1$ $M_\oplus$, we would infer
$s_{\rm max} \sim 10$cm.

In the absence of significant stellar wind, hydrodynamic blowout and
drag are negligible. Whereas radiation blow out remove smallest
grains (with $s < 60 \mu$m), the PR drag causes intermediate-size
($\sim 100 \mu$m) grains to undergo orbit decay on a time scale
($\tau_{\rm PR}$) of 3-50 Myr over this region (between 35-150AU).
Since this time scale is much longer than the time lapse since the
AGB mass loss phase around WD 2226-210, it has negligible effect on
the ring at the center of the Helix nebula. But around mature white
dwarfs (with $\tau_\ast > \tau_{\rm PR}$), it is possible for
\begin{equation}
\tau_\Delta (\Delta = R_R, \tau_{\rm decay} = \tau_{\rm PR}) >
\tau_{\rm syn}
\end{equation}
or equivalently, $\tau_{\rm PR} > (M_\ast / M_p)^{2/3} P_K$. In this
limit, these grains may not be able to bypass the planet's orbit so
that their asymptotic surface density distribution (around older
white dwarfs) may bear the signature of planet's tidal barrier and
have an inner edge (Wyatt {\it et al.} 1999). Nevertheless, one or
more eccentric planets can disable the orbits of these
planetesimals, especially after a substantial loss of $M_\ast$
(Levison {\it et al.} 1994) This process may also provide a supply
of heavy elements to pollute the atmosphere of DZ white dwarfs
(Debes \& Sigurdsson 2002). We will consider that effect elsewhere.

In the case of WD 2226-210 at the center of the Helix nebula, there
may still be a reduced but non negligible stellar wind. Due to
hydrodynamic drag (rather than stellar photons), sub-mm grains are
may be blow away while slightly larger grains may undergo orbital
decay and bypass any embedded planets. In this limit, grain
depletion would lead to transitions in their size distribution. The
inferred values of $s_{\rm max}$ could be reduced by the depletion
of the small particles.

Finally, we consider possible observable consequence of this process.
During the planetary nebula, embedded planets may also accreted gas from
the outflowing stellar wind at a rate
\begin{equation}
{\dot M}_g \simeq \left( {M_p \over M_\ast} \right)^2 \left( {v_K
\over v_{wind}} \right)^4 {\dot M}_\ast.
\end{equation}
The above equation is derived from a modified Bondi accretion
formula. For a Jupiter mass planet, the total accreted gas during
the AGB mass loss phase ($\sim {\dot M}_g \tau_M$) would be a small
fraction ($\sim M_p / M_\ast$) of its own mass. Nevertheless, it
could provide an accretion luminosity (up to $\sim 10^{-4} L_\odot$)
which may be marginally detectable by ALMA against the ring
background.

The hot white dwarf WD 2226-210 at the center of the Helix nebula
provides an intense source of UV radiation. The ionization of
planetary atmosphere can lead to its photoevaporation and mass loss
(Lammer {\it et al.}  2003, Baraffe 2004, 2005, Murray-Clay {\it et
al.} 2009). Lyman $\alpha$ absorbing clouds have been observed
around a close-in planet, HD 209458b (Vidal-Madjar {\it et al.}
2003). Its extension beyond the planet's roche lobe has been
attributed to be a signature of outflow, albeit the inferred mass
loss rate remains uncertain (Ben-Jaffel 2007). The energy flux of
ionizing photons irradiated onto the surface of gas giants located
at $\sim$ 30 AU from WD 2226-210 is 2-3 orders of magnitude larger
than that onto hot Jupiters around solar type stars. Scaling from
the radiative hydrodynamic models for HD 209458b (Murray-Clay {\it
et al.} 2009), we do not expect long-period gas giants to loss
significant fraction of their mass during the AGB mass loss phase.
Nevertheless, photoionizing stellar photons may induce emission of
signature spectral lines on the extended planetary envelope which
may potentially be detectable with high dispersion spectroscopic
observations (we thank Y.-H. Chu for this suggestion).

Mean motion resonant capture of km-size planetesimals by one or more
embedded planets and their subsequent lock-step orbital migration
also lead to their non axisymmetric surface density distribution
(Murray-Clay \& Chiang 2005). In principle, such a structure can be
resolved with sub arc second imaging. However, the observed IR
radiation is due to reprocessed stellar radiation by the fragments
of high-speed collisions. Their recoil velocity may be sufficient to
smear out such a structure. However, multi-wavelength resolved
images may reveal a particle size dependence in the grains radial
distribution.  The detection of very small ($s<s_{\rm blowout}$)
grains at large disk radii and intermediate $s$ ($s_{\rm blowout} <
s < s_{\rm bypass}$) would provide supporting evidence for our
scenario.

\section{Summary and discussion}

In this paper we construct a dynamic model for the post main
sequence evolution of planetary systems. With this model, we find
that as a consequence of mass loss from the central star, km-size
planetesimals can be captured by mean-motion resonance and form a
disk beyond the planets. During the subsequent orbital expansion,
the eccentricity of the resonant planetesimals is excited which
enhances their collision frequency. Highly destructive impacts among
these planetesimals can continuously provide a ring of dust around
the mature star, which can be seen in the FIR and MIR wave band.
This phenomenon can also be extended to other A-F stars, with the
inference that a large fraction of stars may have planet systems and
substantial debris disks during their main sequence epoch. Stars
with higher main sequence mass, greater mass loss rate and lower
velocity of star wind tend to have higher ability to capture the
planetesimals.

We applied this model to account for the dust ring around WD
2226-210 at the center of the Helix nebula. It can also be used to
study other young white dwarfs. For example, we anticipate this
process will occur when the Sun evolve to its AGB mass loss phase,
albeit among of fragmentary dust would be limited and the intensity
of the reprocessed radiation may be weak. Since this process depends
sensitively on the initial heavy element content and the presence of
sufficiently massive planets, we expect there to be large dispersion
in the signatures of dust debris around young white dwarfs.
Nevertheless, we expect to see more dusty rings around planetary
nebulae with A star progenitors because the fractions of stars with
planets and debris disks is observed to increase with stellar mass.

The present work is develpoed for the $\rho_{wind} \propto 1/r^2$ (or $v_{wind}$~=~constant) portion of a stellar wind and thus not applicable to the region very close to the star. In addition, apart from the planetesimals that are captured by the
planet, the fate of other planetesimals also need to be discussed.
These may also play an important role in the evolution of the white
dwarf.

Some observational tests of these models should be possible with present technology or will be possible soon. As the
planetesimals are collected by the giant planet, it is most likely
that the gas giant planet survives the AGB period, and the orbit
changes to several times the original one. If such giant planet
around white dwarf can be detected, it would provide support to our
model. Although there are several failed attempts in these searches
(Debes {\it et al.} 2005a, b, 2007), the upper mass limits set for
the unseen planet is far larger than that needed in our model.

We thank Sverre Aarseth for provide the Hermit integrator, Y.-H.
Chu and an anonymous referee for useful correspondence and helpful
suggestions. This work is
supported by NSFC(10233020), NCET (04-0468), NASA (NNX07A-L13G,
NNX07AI88G, NNX08AM84G), JPL (1270927), and NSF(AST-0908807).

\clearpage

\end{document}